\documentclass[twocolumn]{aastex631}

\usepackage[T1]{fontenc}
\DeclareSymbolFont{UPM}{U}{eur}{m}{n}
\DeclareMathSymbol{\umu}{0}{UPM}{"16}
\let\oldumu=\umu
\renewcommand\umu{\ifmmode\oldumu\else\math{\oldumu}\fi}
\newcommand\micro{\umu}
\renewcommand\micron{\micro$m}
\newcommand\microns{\micron}

\newcommand\methane{\ifmmode{{\rm CH}_{4}}\else{CH$_{4}$}\fi}
\newcommand\water{\ifmmode{{\rm H}_{2}{\rm O}}\else{H$_{2}$O}\fi}
\newcommand\carbdiox{\ifmmode{{\rm CO}_{2}}\else{CO$_{2}$}\fi}
\newcommand\ammonia{\ifmmode{{\rm NH}_{3}}\else{NH$_{3}$}\fi}
\newcommand\acetylene{\ifmmode{{\rm C}_{2}{\rm H}_{2}}
                        \else{C$_{2}$H$_{2}$}\fi}

\usepackage{multirow}
                        
\accepted{July 11, 2022}

\shorttitle{WASP-80}
\shortauthors{Carleo et al.}
\graphicspath{{./}{figures/}}
\begin{document}

\title{The GAPS Programme at TNG XXXIX \footnote{
   Based on observations made with the Italian {\it Telescopio Nazionale Galileo} (TNG) operated by the {\it Fundaci\'on Galileo Galilei} (FGG) of the {\it Istituto Nazionale di Astrofisica} (INAF) at the {\it  Observatorio del Roque de los Muchachos} (La Palma, Canary Islands, Spain).} \\ Multiple molecular species in the atmosphere of the warm giant planet WASP-80\,b unveiled at high resolution with GIANO-B}

\correspondingauthor{Ilaria Carleo}
\email{ilariacarleo.astro@gmail.com}

\author[0000-0002-0810-3747]{Ilaria Carleo}
\affiliation{Astronomy Department, Indiana University, Bloomington, IN 47405-7105, USA}
\affiliation{INAF -- Osservatorio Astronomico di Padova, Vicolo dell'Osservatorio 5, I-35122 Padova, Italy}

\author[0000-0001-7034-7024]{Paolo Giacobbe}
\affiliation{INAF -- Osservatorio Astrofisico di Torino, Via Osservatorio, 20, 10025 Pino Torinese TO, Italy}

\author[0000-0002-1259-2678]{Gloria Guilluy}
\affiliation{INAF -- Osservatorio Astrofisico di Torino, Via Osservatorio, 20, 10025 Pino Torinese TO, Italy}
\affiliation{Observatoire Astronomique de l'Universit\'e de Gen\`eve, Chemin Pegasi 51b, 1290, Versoix, Switzerland}

\author{Patricio E. Cubillos}
\affiliation{INAF -- Osservatorio Astrofisico di Torino, Via Osservatorio, 20, 10025 Pino Torinese TO, Italy}
\affiliation{Space Research Institute, Austrian Academy of Sciences, Schmiedlstraße 6, Graz 8042, Austria}

\author{Aldo S. Bonomo}
\affiliation{INAF -- Osservatorio Astrofisico di Torino, Via Osservatorio, 20, 10025 Pino Torinese TO, Italy}

\author{Alessandro Sozzetti}
\affiliation{INAF -- Osservatorio Astrofisico di Torino, Via Osservatorio, 20, 10025 Pino Torinese TO, Italy}

\author[0000-0002-7704-0153]{Matteo Brogi}
\affiliation{Department of Physics, University of Warwick, CV4 7AL, Coventry, UK}
\affiliation{INAF -- Osservatorio Astrofisico di Torino, Via Osservatorio, 20, 10025 Pino Torinese TO, Italy}
\affiliation{Centre for Exoplanets and Habitability, University of Warwick, CV4 7AL, Coventry, UK}

\author{Siddharth Gandhi}
\affiliation{Department of Physics, University of Warwick, CV4 7AL, Coventry, UK}
\affiliation{Centre for Exoplanets and Habitability, University of Warwick, CV4 7AL, Coventry, UK}
\affiliation{Leiden Observatory, Leiden University, Postbus 9513, 2300 RA Leiden,The Netherlands}

\author[0000-0003-4426-9530]{Luca Fossati}
\affiliation{Space Research Institute, Austrian Academy of Sciences, Schmiedlstraße 6, Graz 8042, Austria}

\author[0000-0002-1923-7740]{Diego Turrini}
\affiliation{INAF -- Osservatorio Astrofisico di Torino, Via Osservatorio, 20, 10025 Pino Torinese TO, Italy}

%%%%%%%%%%%%%%%%%% GAPS-ST, COMMENTI RICEVUTI DA: %%%%%%%%%%%%%%%%%%%%%%%%%%%%
\author{Katia Biazzo}
\affiliation{INAF - Osservatorio Astrofisico di Catania, Via S. Sofia 78, I-95123 Catania, Italy}

\author{Francesco Borsa}
\affiliation{INAF - Osservatorio Astronomico di Brera, Via E. Bianchi 46, 23807 Merate, Italy}

\author[0000-0001-5928-7251]{Antonino F. Lanza}
\affiliation{INAF - Osservatorio Astrofisico di Catania, Via S. Sofia 78, I-95123 Catania, Italy}

\author[0000-0002-6492-2085]{Luca Malavolta}
\affiliation{Dipartimento di Fisica e Astronomia Galileo Galilei, Universit{\'a} di Padova, Vicolo dellOsservatorio 3, I-35122 Padova, Italy}
\affiliation{INAF -- Osservatorio Astronomico di Padova, Vicolo dell'Osservatorio 5, I-35122 Padova, Italy}

\author{Antonio Maggio}
\affiliation{INAF - Osservatorio Astronomico di Palermo, Piazza del Parlamento, 1, I-90134 Palermo, Italy}

\author[0000-0002-9428-8732]{Luigi Mancini}
\affiliation{Department of Physics, University of Rome ``Tor Vergata'', Via
della Ricerca Scientifica 1, I-00133 Rome, Italy}
\affiliation{Max Planck Institute for Astronomy, K\"{o}nigstuhl 17, 69117 --
Heidelberg, Germany}
\affiliation{INAF -- Osservatorio Astrofisico di Torino, Via Osservatorio, 20, 10025 Pino Torinese TO, Italy}

\author{Giusi Micela}
\affiliation{INAF - Osservatorio Astronomico di Palermo, Piazza del Parlamento, 1, I-90134 Palermo, Italy}

\author[0000-0002-1321-8856]{Lorenzo Pino}
\affiliation{INAF - Osservatorio Astrofisico di Arcetri, Largo Enrico Fermi 5, 50125 Firenze, Italy}

\author[0000-0003-1200-0473]{Ennio Poretti}
\affiliation{Fundaci{\'o}n Galileo Galilei-INAF, 
Rambla Jos{\'e} Ana Fernandez P{\'e}rez 7, 38712 Bre{\~n}a Baja, TF, Spain}

\author[0000-0002-8786-2572]{Monica Rainer}
\affiliation{INAF - Osservatorio Astronomico di Brera, Via E. Bianchi 46, 23807 Merate, Italy}

\author[0000-0003-2029-0626]{Gaetano Scandariato}
\affiliation{INAF - Osservatorio Astrofisico di Catania, Via S. Sofia 78, I-95123 Catania, Italy}

\author{Eugenio Schisano}
\affiliation{Institute for Space Astrophysics and Planetology INAF-IAPS, Via Fosso del Cavaliere 100, 00133 Rome, Italy}

%%%%%%%%%%%%%%%%%% TNG + GAPS-BOARD + Seth %%%%%%%%%%%%%%%%%%%%%%%%%%%%%%
\author[0000-0001-5125-6397]{Gloria Andreuzzi}
\affiliation{Fundaci{\'o}n Galileo Galilei-INAF, 
Rambla Jos{\'e} Ana Fernandez P{\'e}rez 7, 38712 Bre{\~n}a Baja, TF, Spain}
\affiliation{INAF - Osservatorio Astronomico di Roma, Via Frascati 33, 00078 Monte Porzio Catone, Italy}

\author[0000-0002-5606-6354]{Andrea Bignamini}
\affiliation{INAF - Osservatorio Astronomico di Trieste, Via Tiepolo 11, 34143 Trieste, Italy}

%\author{Riccardo Claudi}
%\affiliation{INAF -- Osservatorio Astronomico di Padova, Vicolo dell'Osservatorio 5, I-35122 Padova, Italy}

\author[0000-0003-1784-1431]{Rosario Cosentino}
\affiliation{Fundaci{\'o}n Galileo Galilei-INAF, 
Rambla Jos{\'e} Ana Fernandez P{\'e}rez 7, 38712 Bre{\~n}a Baja, TF, Spain}

%\author{Elvira Covino}
%\affiliation{INAF - Osservatorio Astronomico di Capodimonte, Salita Moiariello 16, 80131 Napoli, Italy}

%\author{Silvano Desidera}
%\affiliation{INAF -- Osservatorio Astronomico di Padova, Vicolo dell'Osservatorio 5, I-35122 Padova, Italy}

\author{Aldo Fiorenzano}
\affiliation{Fundaci{\'o}n Galileo Galilei-INAF, 
Rambla Jos{\'e} Ana Fernandez P{\'e}rez 7, 38712 Bre{\~n}a Baja, TF, Spain}

\author{Avet Harutyunyan}
\affiliation{Fundaci{\'o}n Galileo Galilei-INAF, 
Rambla Jos{\'e} Ana Fernandez P{\'e}rez 7, 38712 Bre{\~n}a Baja, TF, Spain}

\author{Emilio Molinari}
\affiliation{INAF Osservatorio Astronomico di Cagliari \& REM, Via della Scienza, 5, I-09047 Selargius CA, Italy}

%\author{Isabella Pagano}
%\affiliation{INAF - Osservatorio Astrofisico di Catania, Via S. Sofia 78, I-95123 Catania, Italy}

\author[0000-0002-5752-6260]{Marco Pedani}
\affiliation{Fundaci{\'o}n Galileo Galilei-INAF, 
Rambla Jos{\'e} Ana Fernandez P{\'e}rez 7, 38712 Bre{\~n}a Baja, TF, Spain}

%\author{Giampaolo Piotto}
%\affiliation{Dipartimento di Fisica e Astronomia Galileo Galilei, Universit{\'a} di Padova, Vicolo dellOsservatorio 3, I-35122 Padova, Italy}

\author[0000-0003-3786-3486]{Seth Redfield}
\affiliation{Astronomy Department and Van Vleck Observatory, Wesleyan University, Middletown, CT 06459, USA}

\author{Hristo Stoev}
\affiliation{Fundaci{\'o}n Galileo Galilei-INAF, 
Rambla Jos{\'e} Ana Fernandez P{\'e}rez 7, 38712 Bre{\~n}a Baja, TF, Spain}



\begin{abstract}
Detections of molecules in the atmosphere of gas giant exoplanets allow us to investigate the physico-chemical properties of the atmospheres. Their inferred chemical composition is used as tracer of planet formation and evolution mechanisms. Currently, an increasing number of detections is showing a possible rich chemistry of the hotter gaseous planets, 
%indicating that such diversity of atomic and molecular constituents may be a common characteristic of these planets, 
but whether this extends to cooler giants is still unknown. We observed four transits of WASP-80\,b, a warm transiting giant planet orbiting a late-K dwarf star with the near-infrared GIANO-B spectrograph installed at the Telescopio Nazionale Galileo and performed high resolution transmission spectroscopy analysis. We report the detection of several molecular species in its atmosphere. Combining the four nights and comparing our transmission spectrum to planetary atmosphere models containing the signature of individual molecules within the cross-correlation framework, we find the presence of H$_{2}$O, CH$_{4}$, NH$_{3}$ and HCN with high significance, tentative detection of CO$_{2}$, and inconclusive results for C$_{2}$H$_{2}$ and CO. 
% We qualitatively interpret these results with... 
A qualitative interpretation of these results, using
physically motivated models, suggests an atmosphere consistent with
solar composition and the presence of disequilibrium chemistry and we therefore recommend the inclusion of the latter in future modelling of sub-1000K planets.

\end{abstract}



\section{Introduction} \label{sec:intro}
%Hot and warm giant planets and transmission spectroscopy are a successful combination to advance our understanding on exoplanets. In fact, the former have relatively easily detectable atmosphere signature; and the latter is a powerful technique which allows to probe the presence of atomic and molecular species in the atmosphere of exoplanets. Given the dependence of the atmospheric scale height $H$ with the equilibrium temperature T$_{eq}$, the mean molecular mass $\mu$ and the surface gravity $g$ ($H$=K$_{b}$T$_{eq}$/$\mu$ $g$), and given the relation between the scale height $H$ and the amplitude of spectral features in transmission $\delta_{\lambda}$ ($\propto$ $R_{\rm P}H/R_{\star}^2$), warmer and gaseous planets with low $g$ and low $\mu$ (hydrogen- and helium- dominated atmospheres) are the ideal targets for atmospheric studies through the transmission spectroscopy technique. Nevertheless, the strength of the atmospheric signature for these candidates is still very small ($\sim$ 10$^{-4}$-10$^{-5}$). 

More and more studies of the atmospheres of hot and warm Jupiters, 
i.e. gaseous giant planets with equilibrium temperatures of $T_{\rm eq} \gtrsim 1000$~K and $T_{\rm eq} \lesssim 1000$~K respectively, 
are yielding important advances in our understanding of the properties of exoplanetary atmospheres and of their possible links to planet formation and migration mechanisms \citep[e.g.,][]{Madhusudhan2019}.
%These hot giant planets have relatively easily detectable atmosphere signatures in both emission and transmission spectroscopy, the latter being a powerful technique to probe the presence of atomic and molecular species at the atmospheric terminator. 
%Indeed, the amplitude of spectral features in transmission spectroscopy is $\propto$ $R_{\rm P}H/R_{\star}^2$, where, 
These hot and warm giant planets are the ideal targets for atmospheric studies through the transmission spectroscopy technique, which allows us to probe the presence of atomic and molecular species at the atmospheric terminator during planetary transits. 
Indeed, the amplitude of spectral features in transmission spectroscopy is %$\delta_{\lambda}$
$\propto$ $R_{\rm P}H/R_{\star}^2$, where $R_{\rm P}$ and $R_{\star}$ are the planet and stellar radii, and $H$=k$_{b}T_{\rm eq}$/$\mu$ $g$ is the atmospheric scale height, with k$_{b}$ the Boltzmann's constant, $\mu$ the mean molecular weight and $g$ the planet surface gravity. Planets with higher $T_{\rm eq}$ and lower $g$ and $\mu$ (hydrogen- and helium-dominated atmospheres) are thus the most favorable for atmospheric studies.

%Given the dependence of the atmospheric scale height $H$ with the equilibrium temperature T$_{eq}$, the mean molecular mass $\mu$ and the surface gravity $g$ ($H$=K$_{b}$T$_{eq}$/$\mu$ $g$), and given the relation between the scale height $H$ and , warmer and gaseous planets with low $g$ and low $\mu$ (hydrogen- and helium- dominated atmospheres) are the ideal targets for atmospheric studies through the transmission spectroscopy technique. Nevertheless, the strength of the atmospheric signature for these candidates is still very small ($\sim$ 10$^{-4}$-10$^{-5}$). 

%Being very close to their host stars, giant planets' atmospheres absorb high-energy radiation, with the consequence of an increasing temperature and subsequent atmospheric expansion. 
%Although the signal of interest -- the effective planet-star radius ratio -- is very small ($\sim$ 1\% for a Jupiter-size planet orbiting a Sun-like star), hot giant transiting planets are the most suitable targets for atmospheric studies through the transmission spectroscopy technique. 

To date most of the insight from the atmospheres of transiting exoplanets comes from low-resolution ($R \sim 200-2000$) spectroscopy (LRS), especially from space thanks to \textit{HST}, both in the optical  and  the  near-InfraRed  (nIR)  wavelength  ranges \citep[e.g.,][]{Sing2016,Mansfield2021}. High-resolution ($R \gtrsim 30000$) spectroscopy (HRS), resolving the molecular absorption bands into thousands of individual lines, has also been proving an effective tool in the investigation of exoplanetary atmospheres \citep[see e.g.][for a review]{Birkby2018}, providing additional/complementary information to the low-resolution data. Indeed, while LRS is sensitive to broad-band absorption features and the level of the spectral continuum relative to the stellar one (which gives information on
the overall transit depth of the planet), HRS is sensitive to the core of the lines, and gives information on the line shape, line Doppler-shift, line-to-line and line-to-continuum contrast. This allows us to investigate higher layers (lower pressures) of the atmospheres, thus possibly studying layers lying above possible aerosol layers \citep{Gandhi2020, Hood2020}. %, which instead obscure the molecule features in the low-resolution spectra. 

%The first nIR ground-based detection with this technique has been presented by \cite{Snellen2010}, revealing absorption lines from carbon monoxide in HD209458\,b.
%As of today, most of the detections have been reported in the atmospheres of tens of hot Jupiters, both in visible and nIR bands, and 

%By analysing nIR high resolution (HR) spectra with a Principal Component Analysis (PCA) approach combined with a dedicated selection of spectral orders, 

While past nIR HRS observations of warm and hot transiting giant exoplanets have detected at most two species, \cite{Giacobbe2021} have recently reported the detection of multiple molecules in the atmosphere of HD209458\,b, revealing a rich chemistry in this hot Jupiter and a Carbon-to-Oxygen ratio (C/O) close to or greater than 1, under the assumption of chemical equilibrium. This estimate of C/O would imply that the planet formed beyond the water condensation front (snowline) at about 2-3 au, and then migrated inward  without substantial accretion of oxygen-rich solids or gas.
%\citep[][and references therein]{Giacobbe2021}.
%and the power of the high-resolution spectroscopy technique.
Whether the rich chemistry found in the atmosphere of HD209458\,b pertains to other hot giant planets and the less studied warm Jupiters is unknown. 

%Recently, warm Jupiters are becoming appealing targets, especially because they are a key missing piece to our understanding of the planetary formation and evolution theories. Warm giant planets origin channels are debated among in-situ, disk migration and high-eccentricity tidal migration. Different theories lead to different observed properties, including the atmospheric composition. In fact, metallicity and elemental chemical abundances, such as the carbon-to-oxygen ratio C/O, can provide important information on the formation region in the protoplanetary disk and evolution path of the planet. For these reasons, the interest in analysing cooler (T$_{eq}$ $\textless$ 1000 K) giant planets at high resolution is rapidly increasing, and the study of their chemistry can add a very important hint to our understanding of warm Jupiters origins. 
   
    %By analyzing nIR spectra from the high-resolution spectrograph GIANO-B \citep{Oliva2006}, \citealt{Giacobbe2021} detected for the first time simultaneously six carbon-, oxygen- and nitrogen-bearing molecules in the atmosphere of one of the best-studied HJs, HD209458b, namely H$_2$O (water), HCN (hydrogen cyanide), CH$_4$ (methane), NH$_3$ (ammonia), CO (carbon monoxide), and C$_2$H$_2$ (acetylene). 

WASP-80\,b is a transiting warm giant planet (T$_{\rm eq}$= 817 K) that orbits a relatively active cool (late-K) dwarf every 3.07 days \citep{Triaud2013}. It has a radius of 0.952 R$_{\rm J}$ and a mass of 0.54 M$_{\rm J}$ \citep[the system parameters are listed in Table \ref{tab:params}]{Triaud2013,Mancini2014,Bonomo2017}. The resulting low density of $\sim$0.8 g/cm$^3$ and the large transit depth of $\sim$3\% make it a very good candidate for transmission spectroscopy. 

Transmission spectroscopy with different datasets and ground-based instruments points to contrasting results. In fact, while \citet{Mancini2014} could not infer, from their multi-color photometry analysis of WASP-80\,b any variation in the planetary radius with wavelength due to large errors in the data, \citet{Kirk2018} found their transmission spectrum best represented by a Rayleigh scattering slope, which indicated the presence of hazes. Moreover, \citet{Sedaghati2017} claimed a detection of the pressure-broadened K I doublet, suggesting a clear and low-metallicity atmosphere for WASP-80\,b, while \citet{Parviainen2018} found the opposite result, showing a flat transmission spectrum, with no significant K I and Na I absorptions. Most recently, \citealt{Fossati2022} did not find any He planetary absorption, possibly indicating a low He abundance in the atmosphere of WASP-80\,b.
From the space-based \textit{HST/WFC3} data, \citet{Tsiaras2018} found no significant presence of water in the atmosphere of WASP-80\,b. Also, \citet{FisherHeng2018} performed a retrieval analysis on the \textit{HST/WFC3} data of WASP-80\,b: although the best-fit model is the grey-cloud model with the water feature (see Fig. 22 in \citealt{FisherHeng2018}), the Bayesian statistics does not favour this model over a flat line, meaning no conclusive retrieved atmospheric properties can be reported.

In this letter, we present the analysis of the transmission spectrum of WASP-80\,b using the HR GIANO-B data and the same technique as in \citet{Giacobbe2021}.  The observations and data reduction are described in Sec. \ref{sec:obs}, the resulting detections of molecules are presented in Sec. \ref{sec:detections}, which contains the central findings of this work. While this letter is mainly focused on the detection of multiple species, 
%Without the ambition to quantify chemical abundances and/or distinguish equilibrium and disequilibrium conditions, 
we give possible qualitative interpretations in Sec. \ref{sec:model} and, finally, the conclusions are given in Sec. \ref{sec:concl}.

\begin{table}[htbp]
\centering
\caption{\label{tab:params} Parameters of the WASP-80 system.}
\begin{tabular}{lcl}
\noalign{\smallskip}
\hline
\noalign{\smallskip}
Parameter &  Value & Reference  \\
\noalign{\smallskip}
\hline
\noalign{\smallskip}
$M_{*}$ &  0.570$\pm$0.050 M$_{\odot}$ & \citealt{Triaud2013}\\
$R_{*}$ &  0.571$\pm$0.016 R$_{\odot}$ & \citealt{Triaud2013}\\
$T_{\rm eff}$ & 4150$\pm$100 K & \citealt{Triaud2013}\\
$\rm [Fe/H]$ & -0.14$\pm$0.16 dex & \citealt{Triaud2013} \\
%logg$_*$ & 4.663$^{+0.015}_{-0.016}$ (cgs) &\citealt{Triaud2015}\\
$v\,\sin{i}$ &  3.55$\pm$0.33 km\,s$^{-1}$ &\citealt{Triaud2013}\\
$\log{\rm g}_*$ & 4.689$^{+0.012}_{-0.013}$ (cgs) &\citealt{Triaud2013}\\
$\log{\rm R}^\prime_{\rm HK}$ & -4.04$\pm$0.02  & \citealt{Fossati2022} \\
$P_{\rm orb}$ & 3.06785234$^{+0.00000083}_{-0.00000079}$ days & \citealt{Triaud2015}\\
$M_{\rm P}$ & 0.540$^{+0.035}_{-0.036}$ M$_{\rm J}$ & \citealt{Bonomo2017}\\
$R_{\rm P}$ & 0.952$^{+0.026}_{-0.027}$ R$_{\rm J}$ & \citealt{Triaud2013}\\
$K_{\rm P}$ & 122$\pm$4 km\,s$^{-1}$ & This work\\
$T_{\rm eq}$ & 817$\pm$20 K & This work\footnote{We calculate the equilibrium temperature using the Equation 1 in \citet{Lopez2007}, which assumes the Bond albedo equal to zero and redistribution factor f=1/4.}\\
%T$_{\rm eq}$ & 825$\pm$19 K & \citealt{Triaud2015}\\
%logg$_P$ &  &\\
$\rho_{\rm P}$ & 0.776$^{+0.088}_{-0.078}$ g cm$^3$& \citealt{Bonomo2017}\\
a & 0.03427$^{+0.00096}_{-0.00100}$ AU & \citealt{Bonomo2017}\\
%e & 0.0020$^{+0.0100}_{-0.0020}$ & \citealt{Triaud2015}\\
e & $< 0.020$ & \citealt{Bonomo2017}\\
\noalign{\smallskip}
\hline
\noalign{\smallskip}
\end{tabular}
\end{table}

\section{Observations and Data Reduction} \label{sec:obs}
We observed four transits of WASP-80\,b on 2019/08/09, 2019/09/21, 2020/06/26 and 2020/09/17 in GIARPS mode \citep{Claudi2017}, using GIANO-B \citep{Oliva2006, Carleo2018} and HARPS-N \citep{Cosentino2012,Cosentino2014} simultaneously, at the Telescopio Nazionale Galileo (TNG).  While \citealt{Fossati2022} presented an analysis of the optical portion of these observations together with a search for He~$\small{I}$, in this work we mainly exploit the nIR GIANO-B data, in order to search for molecules in the atmosphere of this planet. These observations are part of the GAPS2\footnote{https://theglobalarchitectureofplanetarysystems.wordpress.com/} programme, aimed at exploring the diversity of planetary systems via the detection of planets around young stars \citep[e.g.,][]{Carleo2020yo, Damasso2020}, the search for inner small planetary companions to outer long-period giants \citep[e.g.,][]{Barbato2020}, and the observation and characterization of planetary atmospheres \citep[e.g., Guilluy et al. in prep.,][]{Borsa2019,Pino2020,Guilluy2020,Giacobbe2021}. %Our transit observations were performed in GIARPS mode \citep{Claudi2017}, using GIANO-B \citep{Oliva2006, Carleo2018} and HARPS-N \citep{Cosentino2012,Cosentino2014} simultaneously, at the Telescopio Nazionale Galileo (TNG). 
%In this work, we mainly exploit the nIR GIANO-B data, in order to search for molecules in the atmosphere of this planet. GIANO-B covers a wavelength range of 0.95 - 2.45 $\mu$m split in fifty orders with a resolving power of R$\sim$50000. 
GIANO-B covers a wavelength range of 0.95 - 2.45 $\mu$m split in fifty orders with a resolving power of R$\sim$50000. The spectra were reduced with the offline GOFIO pipeline \citep{Rainer2018}. The first three nights present a median signal-to-noise ratio (S/N) of $\sim$ 30 (with a maximum of $\sim$ 60), whereas the forth night is characterized by a median S/N $\sim$ 20 and a max S/N of $\sim$ 40 (see \citealt{Fossati2022} for more details on the observations, observing log, and data reduction).

\section{Transmission spectroscopy and search for molecules}\label{sec:detections}
%The main focus of this work is the detection of multiple molecular species in the atmosphere of WASP-80\,b. Therefore, these results are not intended to put tight constraints to the chemical abundances or the distinction between thermodynamic equilibrium and non-equilibrium scenarios, and we only give a qualitative interpretation in Sec. \ref{sec:model}. 

\paragraph{\textbf{Cross-Correlation analysis}} For the transmission spectroscopy analysis, we followed the recipe by \citealt{Giacobbe2021}. Briefly, {\it a)} we wavelength-calibrated and aligned all the spectra to the observer's rest frame by using the Earth's absorption lines (telluric) as reference;
{\it b)} we removed the quasi-stationary (in wavelength) spectral components - not only tellurics but also the star at first order - using our custom Principal Component Analysis (PCA) approach; 
%{\it c)} we performed an optimal selection of the spectral orders for each molecule and for each night;
{\it c)} we performed an optimal selection of the spectral orders, for each molecule and for each night, to discard the orders that do not contain enough signal (molecular lines) and/or are strongly contaminated by telluric and stellar lines;
%, by injecting the models in the real data and recovering the signal from the synthetic spectra. The order selection in this way maximizes the detections, assuming the S/N constant within one order and among the orders; 
{\it d)} we applied the cross-correlation (CC) technique between the observed data and the atmospheric models (described below) to investigate the presence of H$_{2}$O, CH$_{4}$, HCN, NH$_{3}$, CO, CO$_{2}$, C$_{2}$H$_{2}$. The cross-correlation function (CCF) is computed over a range of radial velocities between -252 and 252 km\,s$^{-1}$ in steps of 3 km\,s$^{-1}$ and for each molecule, spectral order, night and exposure, and then co-added over all selected orders and nights. Even if we know with relative precision the theoretical planet's orbit radial velocity semi-amplitude K$_{\rm P}$ = 122$\pm$4 km\,s$^{-1}$, we explore a range of K$_{\rm P}$ values between 0 and 200 km\,s$^{-1}$ with steps of 3 km\,s$^{-1}$, in order to explore spurious detections near the expected K$_{\rm P}$ and to account for the uncertainty on K$_{\rm P}$, 
%in order to assess that there is not any spurious detection, and to account for uncertainties on the velocity value given in the literature (a few km/s), 
as well as for dynamical effects of atmospheric winds. The resolution of 3 km/s was chosen to be equal to the half width half maximum of the instrumental profile for R=50,000.
%\textcolor{red}{The atmospheric models for all molecules are generated, separately, with an isothermal profile and constant volume mixing ratio (VMR) for the seven molecules under investigation, and are convolved with the GIANO-B instrumental profile, approximated with a Gaussian function. The temperatures spans between 1,000 ad 1,500 K, and the VMRs between 10$^{-5}$ and 10$^{-2}$. (Sid and Patricio, add references). }

For each molecule we generated atmospheric transmission spectra using isothermal pressure/temperature (P/T) profiles using GENESIS \citep{Gandhi2017, Giacobbe2021}. The choice of isothermal P/T profiles is guided by the fact that in transmission spectroscopy a change in temperature acts to change the planet scale height with altitude, leading in a change on the overall strength of the spectral lines. Thus, changing the shape of the P/T profile from isothermal to a more complex profile would not significantly change the chemistry in the atmosphere and the  molecular detections.

The models span a range of pressure-temperature profiles ([100, 10$^{-8}$] bar, [-200, 200] K around T$_{\rm eq}$) and volume mixing ratios (VMR, [10$^{-12}$, 10$^{-1}$]) for each species and are calculated at a constant wavenumber spacing of 0.01\,cm$^{-1}$ in the 0.9-2.6~$\mu m$ wavelength range. We then convolved the spectra to the instrumental resolution of GIANO assuming a Gaussian profile with a FWHM$\sim$5.4 km\,s$^{-1}$, which corresponds to a spectral resolving power of R$\sim$50000. We adopted the latest and most suitable line lists for high resolution spectroscopy \citep{Gandhi2020}, with CH$_4$ and CO from the HITEMP database \citep{Rothman2010}, H$_2$O, NH$_3$, HCN and C$_2$H$_2$ from ExoMol \citep{Tennyson2016}, and CO$_2$ from Ames \citep{Huang2013, Huang2017}; see Table \ref{tab:detections} for a full list of references for each molecular line list. These are broadened by the pressure and temperature into a Voigt profile according to their H$_2$ and He pressure-broadening coefficients. We additionally include collision induced absorption from H$_2$-H$_2$ and H$_2$-He interactions \citep{Richard2012} as a source of continuum opacity.

%We performed this analysis for each single night, and, in order to increase the final S/N, we sum the four nights up by co-adding the CCFs from the selected orders in phase over all the nights. 
Since the CC is marginally dependent on VMR \citep{Gandhi2020} when considering single species and no clouds, we fixed this parameter and used the highest VMR available for each species for the entire analysis. In particular, VMR=10$^{-1}$ for H$_{2}$O and CO, VMR=10$^{-2}$ for CH$_{4}$, NCH and NH$_{3}$, VMR=10$^{-3}$ for C$_{2}$H$_{2}$ and CO$_{2}$. 
The detection significance is calculated through a Welch $t$-test \citep{Welch1947}, which consists of creating two distributions of cross-correlation values, one in-trail within $\pm$3~km\,s$^{-1}$ around the expected planet's RV, and one out-of-trail considering the values $\pm$25~km\,s$^{-1}$ away from the planet's RV. The null hypothesis is when the two samples have the same mean. The rejection of the null hypothesis represents the significance of the detection. We claim a detection if $\sigma\,\geqslant$\,4.

\begin{figure*}
\centering
\includegraphics[width=1.0\linewidth, trim=4cm 0cm 0cm 0cm,clip]{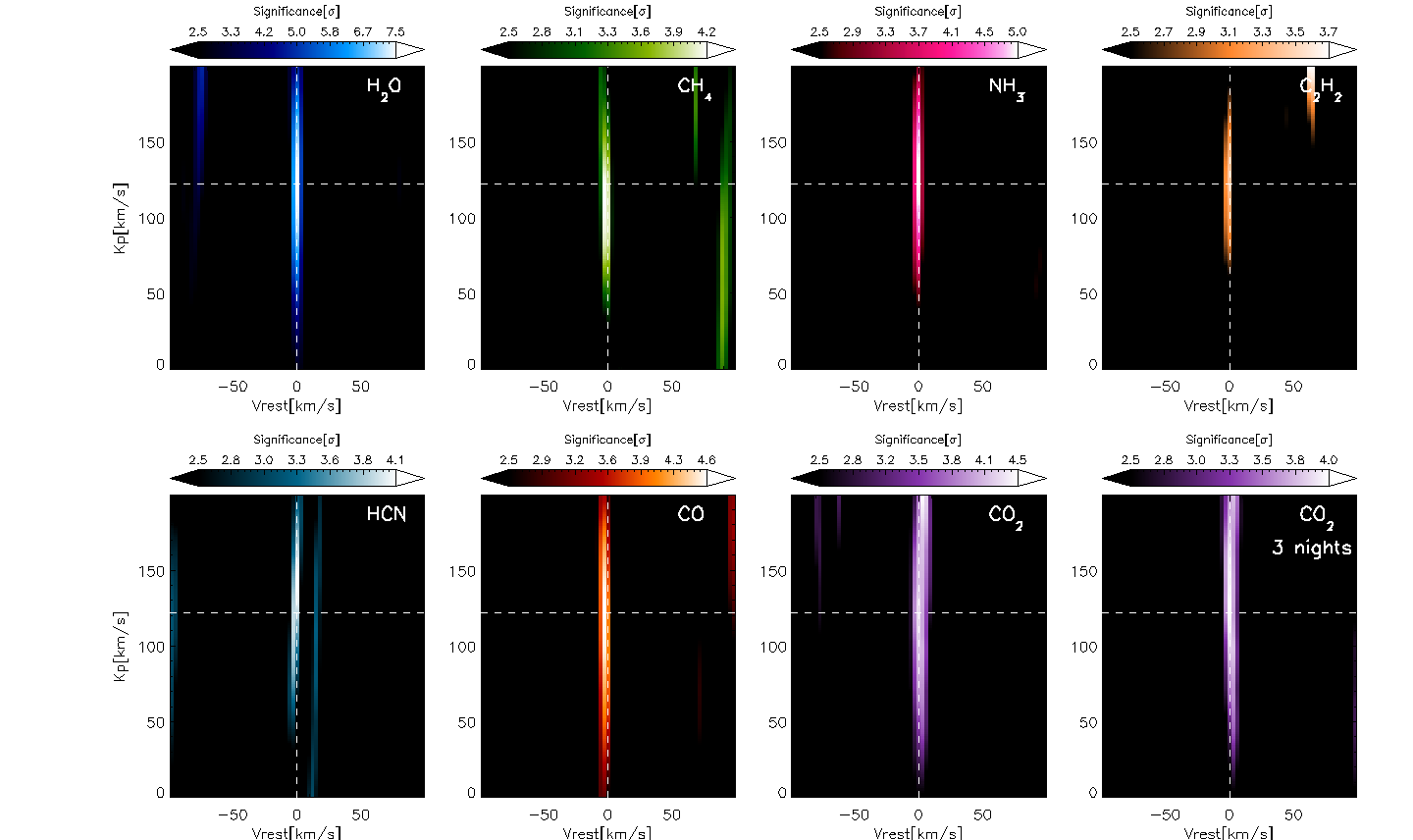}
\caption{\label{fig:detectionssigma} Significance maps of the cross-correlation between the GIANO-B spectra and the isothermal atmospheric models, as a function of the planet’s maximum radial velocity (K$_{\rm P}$) and the planet’s rest-frame velocity (v$_{\rm rest}$). For a better visualization, the maps do not show any signal $\textless$2.5$\sigma$. The dashed white lines indicate the known velocity for WASP-80\,b (K$_{\rm P}$=122~km\,s$^{-1}$, v$_{\rm rest}$=0~km\,s$^{-1}$). The CO$_2$ map is also shown in the case of co-adding only the three nights (not including the third night). See discussion in the text (Sec. \ref{sec:detections}) and Table \ref{tab:detections}. } 
\end{figure*}

From the CC analysis, we detect five out of seven tested species, namely H$_{2}$O (7.5$\sigma$), CH$_{4}$ (4.2$\sigma$), NH$_{3}$ (5$\sigma$), HCN (4.1$\sigma$), CO (4.6$\sigma$). The results are displayed in Table \ref{tab:detections} and the significance maps as a function of the rest-frame velocity v$_{\rm rest}$ and the planetary semi-amplitude K$_{\rm P}$ are shown in Fig. \ref{fig:detectionssigma}. 
%\textcolor{blue}{\textbf{S/N maps??? }}
We find a tentative evidence for C$_{2}$H$_{2}$, since the most significant peak in the map is located at v$_{\rm rest}$=63 km s$^{-1}$ with 3.7$\sigma$,
%3.65
but the second significant peak is at the planetary position with a slightly lower $\sigma$=3.4.
%3.39
%summing up only the first three nights we can detect a signal with a significance of 3.7$\sigma$, but after adding the forth night a spurious peak in the map becomes more significant, even though the planetary signal is still evident and its significance at the nominal position differs much less than 1$\sigma$ from the peak. 
As for CO$_{2}$, the most significant peak (4.45$\sigma$) is at the upper limit of the K$_{\rm P}$ range, even though a signal at the nominal (K$_{\rm P}$, v$_{\rm rest}$) position is visible, though at a slightly lower significance (4.28$\sigma$) than the main peak. Since CO$_2$ can be strongly affected by the tellurics and, during the third night (26 June 2020) the tellurics position falls close to $v_{\rm rest}$ = 0, we performed the analysis for CO$_2$ not including this night, and we found a 4.01$\sigma$ detection at the planetary position  (see Table \ref{tab:detections}), confirming that the result might be affected by telluric contamination. 

The H$_{2}$O signal is so strong that we can detect it in each night separately. Conversely, the rest of the detected molecules have weaker signals and are not always detected in each single transit, but are solidly detected by co-adding the four transits. 
%Moreover, although we do not find CH$_{4}$ in the co-added transit, we can clearly detect it only in the third night at a 4.4$\sigma$. 
%Finally, CO$_{2}$ is not detected in any of the transits. 
These results demonstrate the advantage of the multi-transit strategy when searching for molecular species in planetary atmospheres. 

%The first night (2019/08/09) presents a median SNR of $\sim 25$ and a dome humidity of $\sim$80\% over night, whereas the second night (2019/09/21) is characterized by a median SNR $\sim 30$ and a dome humidity of $\sim$60\% with a big slope (from $\sim$10\% to $\sim$60\%) at the beginning of the observations. 
%Summarizing, the high humidity is reflected in the presence of correlated noise at low $K_p$, in both nights.

%When we sum the two nights, taking advantage of the wavelength shift due to the barycentric velocity of $\sim$ 17 km/s between the nights, we are able to mitigate the presence of the correlated noise.

\begin{table*}[htbp]
\begin{center}
\caption{\label{tab:detections} Summary of the results for each molecule for both Cross-Correlation and likelihood frameworks. }
    \begin{tabular}{l l| c c c | c c c c | c c l}
     \hline
    \noalign{\smallskip} 
        &    &  \multicolumn{3}{c}{\bf Cross-Correlation}      & \multicolumn{4}{|c}{\bf Likelihood} & \multicolumn{3}{|c}{\bf 
            Detection status}\\ 
    \hline
    \noalign{\smallskip}
\multirow{2}{*}{Molecule}  &  \multirow{2}{*}{Database} & v$_{\rm rest}$ & K$_{\rm P}$ & Significance & v$_{\rm rest}$ & K$_{\rm P}$ &  Significance & $\log{S}$ & CC & LH & \\ 
          &   & [km s$^{-1}$] &    [km s$^{-1}$]   & $\sigma$ & [km s$^{-1}$]   &  [km s$^{-1}$] & $\sigma$ &  &  & &\\
\noalign{\smallskip}
\hline
\noalign{\smallskip}
%H$_{2}$O & Exomol & 0.0 & 123.0$\pm$43.5 & 7.52               & 0.0  & 116$^{+15}_{-14}$  & 10.03 & 0.1  & $\checkmark$ & $\checkmark$ & Detected\\
H$_{2}$O & Exomol & 0.0 & 123.0$\pm$43.5 & 7.52               & 0.0  & 116$^{+15}_{-14}$  & 9.94 & 0.1  & $\checkmark$ & $\checkmark$ & Detected\\
%CH$_{4}$& HITEMP  & -3.0 & 115.5$^{+64.5}_{-61.5}$  & 4.17     & -1.0 & 124$^{+34}_{-38}$  & 4.07 & 0.1  & $\checkmark$ & $\checkmark$& Detected\\
CH$_{4}$& HITEMP  & -3.0 & 115.5$^{+64.5}_{-61.5}$  & 4.17     & -1.0 & 124$^{+34}_{-38}$  & 4.08 & 0.1  & $\checkmark$ & $\checkmark$& Detected\\
%NH$_{3}$& Exomol  &  0.0 & 121.5$^{+45.0}_{-55.5}$ & 4.96        & 0.0  & 119$\pm$18 & 8.02 & 0.1   & $\checkmark$ & $\checkmark$& Detected\\
NH$_{3}$& Exomol  &  0.0 & 121.5$^{+45.0}_{-55.5}$ & 4.96        & 0.0  & 119$\pm$18 & 7.63 & 0.1   & $\checkmark$ & $\checkmark$& Detected\\
%C$_{2}$H$_{2}$ (3n)\footnote{First 3 nights} & 0.0 & 126.0$^{+67.5}_{-58.5}$ & 3.71 & 2.0  & 189$^{+11}_{-42}$& 3.36 & 0.0      &  $\sim$ & $\sim$\\
%C$_{2}$H$_{2}$& Exomol   & 0.0 &  129.0$^{+70.5}_{-61.5}$ & 3.39 \footnote{The most significant peak in the CC map is at v$_{\rm rest}$=63 km s$^{-1}$ with a $\sigma$=3.65 (see Fig. \ref{fig:detectionssigma}).} & 0.0  & 155$^{+40}_{-56}$ & 2.62 & -0.1 & $\sim$  & $\times$ & Inconclusive\\
%C$_{2}$H$_{2}$ (4n)\footnote{4 nights} & 63.0 &  199.5$^{+33.0}_{-122.0}$ & 3.65  & 0.0  & 155$^{+40}_{-56}$ & 2.62 & -0.1 & $\times$  & $\sim$ \\
C$_{2}$H$_{2}$& Exomol   & 0.0 &  129.0$^{+70.5}_{-61.5}$ & 3.39 \footnote{The most significant peak in the CC map is at v$_{\rm rest}$=63 km s$^{-1}$ with a $\sigma$=3.65 (see Fig. \ref{fig:detectionssigma}).} & -69.0  & 15$^{+48}_{-15}$ & 3.88 & 0.1 & $\sim$  & $\times$ & Inconclusive\\
%HCN & Exomol & 0.0 & 142.5$^{+57.0}_{-55.5}$ & 4.11            & 0.0  & 134$\pm$30 & 3.38 & -0.5 &$\checkmark$ & $\sim$ & Tentative\\
HCN & Exomol & 0.0 & 142.5$^{+57.0}_{-55.5}$ & 4.11            & 0.0  & 134$\pm$30 & 4.30 & -0.5 &$\checkmark$ & $\checkmark$ & Detected\\
%CO& HITEMP & -3.0 & 124.5$^{+93.0}_{-64.5}$  & 4.62           & -3.0 & 74$^{+25}_{-24}$ & 6.15 & 0.3 & $\checkmark$ & $\times$ & Inconclusive\\
CO& HITEMP & -3.0 & 124.5$^{+93.0}_{-64.5}$  & 4.62           & -3.0 & 74$^{+25}_{-24}$ & 6.33\footnote{Peak is at an inconsistent K$_{\rm P}$} & 0.3 & $\checkmark$ & $\times$ & Inconclusive\\
%CO$_{2}$ (4n)\footnote{4 nights.}& Ames & 3.0 & 199.5$^{+163.5}_{-8.5}$&      4.45       & 2.0  & 64$^{+18}_{-16}$ &  7.36  & 0.2 & $\times$ & $\times$& \multirow{2}{*}{Tentative}\\
%CO$_{2}$ (3n)\footnote{Third night (26 June 2020) not included.}& Ames & 0.0 & 126$\pm$68.5 &  4.01  & -1.0  & 147$^{+18}_{-19}$ &  7.49  & 0.2 & $\checkmark$ & $\checkmark$\\
CO$_{2}$ (4n)\footnote{4 nights.}& Ames & 3.0 & 199.5$^{+163.5}_{-8.5}$&      4.45       & 2.0  & 64$^{+18}_{-16}$ &  8.50$^{\rm b}$  & 0.2 & $\times$ & $\times$& \multirow{2}{*}{Tentative}\\
CO$_{2}$ (3n)\footnote{Third night (26 June 2020) not included.}& Ames & 0.0 & 126$\pm$68.5 &  4.01  & -1.0  & 147$^{+18}_{-19}$ &  7.09  & 0.2 & $\checkmark$ & $\checkmark$\\

\noalign{\smallskip}
\hline
\noalign{\smallskip}
\end{tabular}
\end{center}
References for each molecule:   H$_2$O \citep{Polyansky2018}, CH$_4$ \citep{Hargreaves2020}, NH$_3$ \citep{Coles2019}, C$_2$H$_2$ \citep{Chubb2020}, HCN \citep{Harris2006, Barber2014}, CO \citep{LiEtal2015apjsCOlineList}, CO$_2$ \citep{Huang2013, Huang2017}.
\end{table*}

%\begin{table}[htbp]
%\centering
%\caption{\label{tab:detectionssigma} Detections (significance).}
%\begin{tabular}{lcccl}
%\noalign{\smallskip}
%\hline
%\noalign{\smallskip}
%Molecule &  v$_{\rm sys}$ & K$_{\rm P}$ & Significance & Notes\\
%\noalign{\smallskip}
% &  [km s$^{-1}$] & [km s$^{-1}$] & $\sigma$ & \\
%\noalign{\smallskip}
%\hline
%\noalign{\smallskip}
%H$_{2}$O &  0 & 123.0 & 10.45 &\\
%CH$_{4}$ & -1.5 & 102  & 5.85 &\\
%NH$_{3}$ &  0 & 130.5 & 6.89 &\\
%HCN & -1.5 & 129 & 5.84 &\\
%CO & -3.0 & 121.5  & 5.56 &\\
%C$_{2}$H$_{2}$ & 1.5 & 139.5 & 4.93 & First 3 nights\\
%C$_{2}$H$_{2}$ & 64.5 &  199.5 &  & 4 nights\\
%CO$_{2}$ & 4.5 & 199.5 & 5.27 &\\
%\noalign{\smallskip}
%\hline
%\noalign{\smallskip}
%\end{tabular}
%\end{table}

%\begin{figure*}
%\centering
%\includegraphics[width=0.95\linewidth]{res_SNR_n_OK_4notti.png}
%\caption{\label{fig:detectionssnr} Detections (S/N).}
%\end{figure*}

\paragraph{\textbf{Likelihood analysis}} 
After performing the cross-correlation approach, we convert the CCFs into likelihood (LH) values \citep{Brogi2019}. Briefly, the log-likelihood function is defined as in Eq. 9 of \citet{Brogi2019}:

\begin{equation}
    \log{L} = -\frac{N}{2}\log{[s_f^2 - 2R(s) + s_g^2]},
\end{equation}

where N is the number of spectral channels, s$_f^2$ is the variance of the data, s$_g^2$ the variance of the model and $R(s)$ the cross-covariance between the data and the model with $s$ being a bin/wavelength shift. This function is computed for each order and each spectrum. The final log-likelihood value is the sum of all log-likelihood functions for each night. As in \citet{Giacobbe2021}, an additional free parameter, namely the line-intensity scaling factor $S$, is introduced in this framework. If the model (i.e. the strength of the spectral lines) perfectly matches the data,  $S$ will be 1, that means $\log{ S}$=0. Unlike in the CC (which is a normalized quantity), the line depth is important in the LH framework and the scaling factor allows us to approximate the continuum of the model as well as to take into account the effects from other species.  In this framework, we calculate the significance by comparing the maximum value of the likelihood to the mean LH value in the map used as baseline.

\begin{figure*}
\centering
\includegraphics[width=0.95\linewidth]{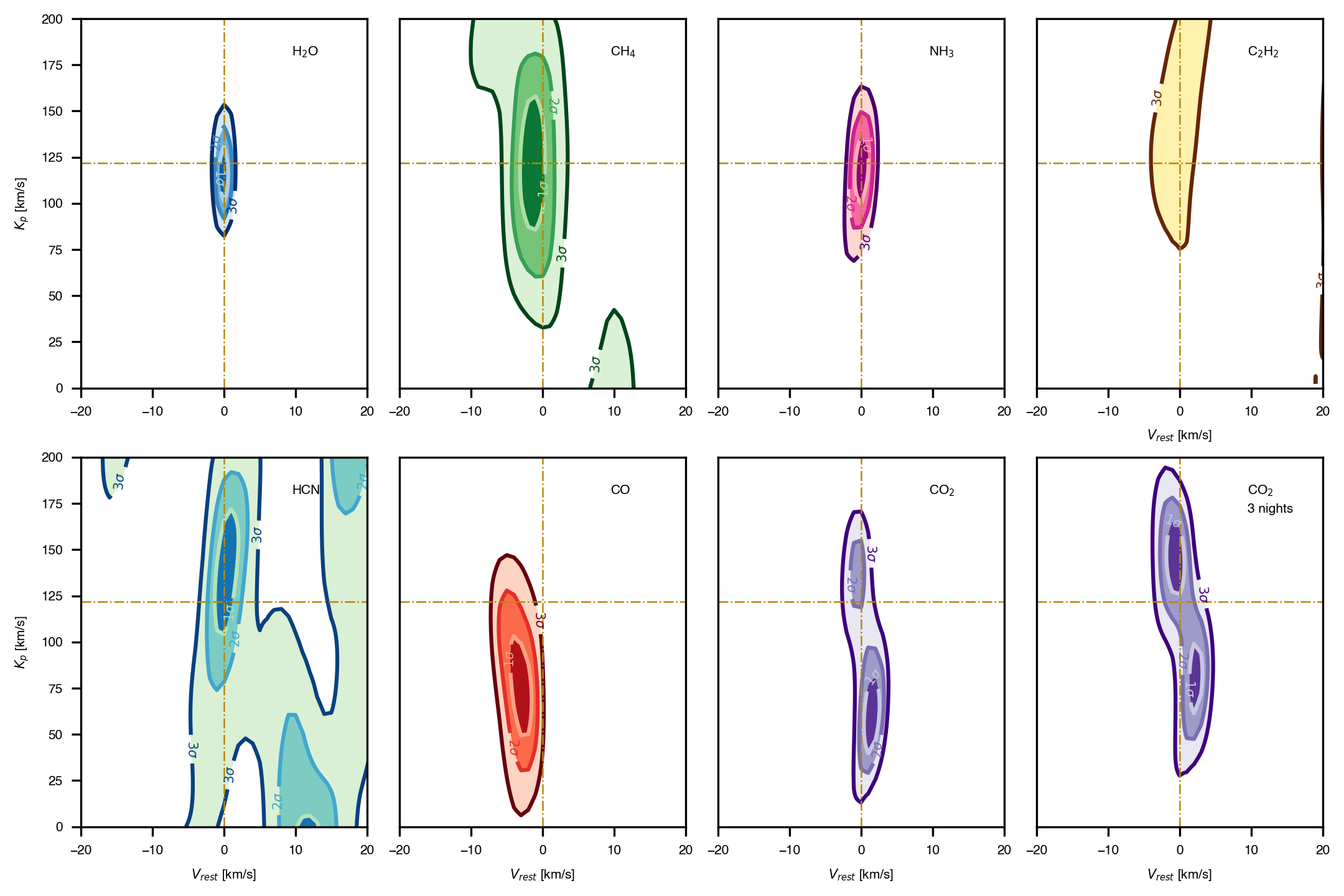}
\caption{\label{fig:detectionslike} Likelihood confidence intervals maps for each investigated molecule as a function of K$_{\rm P}$ and v$_{\rm rest}$ and at the best-fit scaling factor. The dashed lines represent the nominal planetary velocities. As for the CC significance maps, the CO$_2$ likelihood map is displayed in the case of co-adding only three nights (not including the third night). As regards C$_2$H$_2$, the likelihood intervals map does not significantly change by co-adding 3 or 4 nights. }
\end{figure*}

The likelihood function is computed for each order, each observed spectrum and each night on a grid of K$_{\rm P}$ ([0, 200] km\,s$^{-1}$ in steps of 1~km\,s$^{-1}$), v$_{rest}$ ([-99,  99]\,km\,s$^{-1}$ in steps of 1~km\,s$^{-1}$) and $\log{ S}$ ([-1.1, 1.1] in steps of 0.1). Table \ref{tab:detections} reports the v$_{\rm rest}$, K$_{\rm P}$ and $\log{ S}$ values, while Fig. \ref{fig:detectionslike} shows the log-likelihood confidence interval maps for each molecule maximized at the best-fit scaling factor for the model, with a zoomed v$_{rest}$ interval between [-20,  20]\,km\,s$^{-1}$ for a better readability. The original version of this map is shown in Appendix \ref{app:B}. Since in our analysis we used fixed VMRs and single-species model (meaning that the line contrast is deeper than in a mixed model), the log${ S}$ values are only indicative and cannot be properly interpreted.

In this likelihood framework, we confirm four out of five molecules detected with the CC, namely H$_2$O, CH$_4$, NH$_3$ and HCN. 
%even though the latter with a significance lower than 4$\sigma$ (3.38$\sigma$). 
%We also find the signal of C$_2$H$_2$ at the expected (K$_{\rm P}$, v$_{\rm rest}$), but, having a low significance of 2.62$\sigma$, we do not claim its detection. 
As in the case of the CC approach, CO$_2$ is not detected in the 4-night case, showing a significant peak at low K$_{\rm P}$ values, even though a slightly less significant peak (4.28$\sigma$) is present in the likelihood map as well. Furthermore, when we exclude the third night, the peak around the planetary K$_{\rm P}$ becomes the most significant in the map (see Fig. \ref{fig:detectionslike}). 

Finally, the CO signal shows a low value of K$_{\rm P}$ in the likelihood map, which is not compatible with the nominal K$_{\rm P}$
%only at the upper limit of the 2$\sigma$ contour \textcolor{red}{(ma dalla figura non è a piu'\ di tre sigma? Direi che è incompatibile con il Kp plamnetario)} 
(see Fig. \ref{fig:detectionslike}). 
This low value can be due to the contamination of the stellar CO lines, which is quasi-stationary in wavelength. In fact, because of its rotation, the stellar absorption lines are distorted by the Rossiter-McLaughlin (RM) effect and, during the transit, this effect can cause spurious signals at different transit phases, and finally in the detection maps, being of the same order of magnitude of the planetary signal. Unlike other molecules, the RM effect can heavily affect the CO detection \citep{Brogi2016}. In order to investigate this contamination, we first calculate the theoretical expected $K_{\rm P}$ for the stellar atmosphere as $v\,\sin{i}$/sin(2$\pi\phi_{1}$), where $\phi_{1}$ is the phase at transit ingress (0.014 for WASP-80). We find that the expected K$_{\rm P}$ corresponding to the stellar CO is 40.4 km s$^{-1}$, which is lower than the K$_{\rm P}$ value for the LH peak. On the other hand, it is possible that our LH result might be a combination of the stellar and planetary signal, which would appear at intermediate K$_{\rm P}$ values. As an additional test, we generated a synthetic stellar spectrum with the effective temperature, $\log{\rm g}_*$ and metallicity values of WASP-80, from the PHOENIX spectral library \citep{Husser2013}. We convolved this spectrum to the GIANO-B resolution and calculated the CCFs and the LH using the PHOENIX spectrum to mask the stellar lines in our observed spectra. This operation did not improve the previous results, thus resulting inconclusive.

The difference between the CC and LH maps reflects the fact that, while the CC analysis is mostly sensitive to the line position and relative (i.e. line-to-line) amplitude, the LH analysis is additionally sensitive to the line shape and the line-to-wing contrast ratio. Therefore, a model that matches well the line position in CC might still be penalised in the LH analysis if the line shape/amplitude is mismatched. This is also a strong motivation for presenting both the CC and LH analyses even in the absence of a full atmospheric retrieval. While the LH is advantageous in the amount of information extracted from the spectra, it is also more demanding in terms of the accuracy of the modelling. On the other hand, the CC analysis is probably better at detecting species even with imperfect templates, which is valuable at the exploratory stage.

%\begin{figure}
%\centering
%\includegraphics[width=0.95\linewidth]{Schermata 2021-11-09 alle 10.27.22.png}
%\caption{\label{fig:cotest} CO test: masking stellar lines (right panel) and no masking (left panel).}
%\end{figure}

\section{Atmospheric Modeling}
\label{sec:model}

\begin{figure*}[thb!]
\centering
\includegraphics[width=\linewidth]{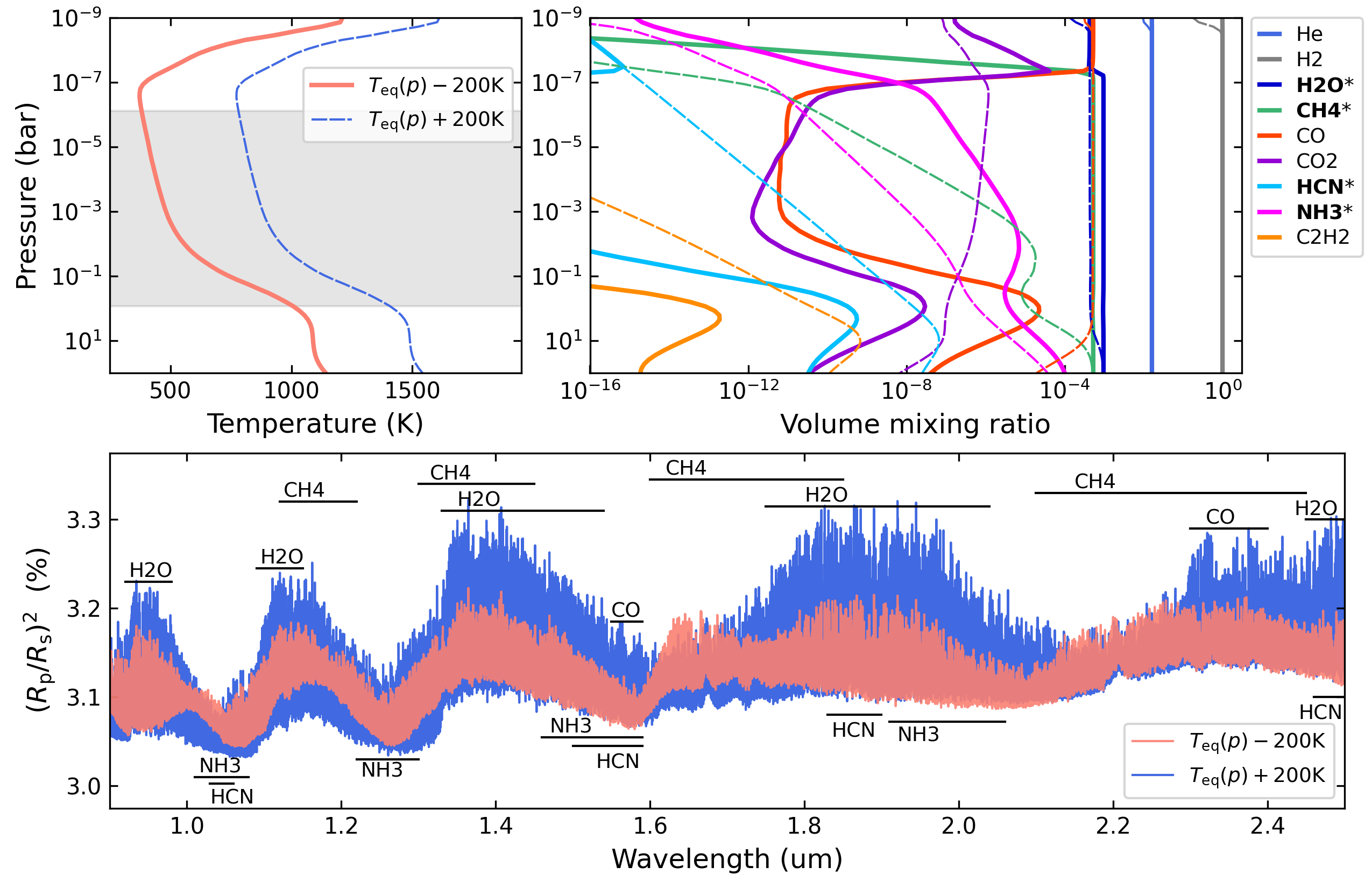}
\caption{Sample atmospheric models for WASP-80\,b in radiative and
  thermochemical equilibrium for a solar elemental composition.  The
  top left panel shows the equilibrium temperature profile shifted by
  $\pm 200$~K, to simulate a low-temperature case (thick solid red
  line) and a high-temperature case (thin dashed blue line). The shaded area denotes the pressure range probed by the GIANO-B transmission observations for these models.
  The top right panel shows the equilibrium composition
  for the two temperature profiles at low (thick solid lines) and high
  temperature (thin dashed lines), color coded for each relevant species (see legend, with the asterisks indicating the firmly detected species).  The bottom panel shows the transmission spectra over
  the GIANO-B wavelength range, for the low-temperature case (red) and
  high-temperature model (blue).  The horizontal black lines denote the
  wavelength ranges where each of the labeled species are expected to significantly shape the transmission spectrum of WASP-80\,b.}
\label{fig:equilibrium_physics}
\end{figure*}

Given the rich set of molecular detections by GIANO-B on WASP-80\,b, we
can make rough estimates of the physical properties of the planet being
guided by the observations, even without quantitative abundance estimates.  In fact, WASP-80\,b sits at a
particularly favorably location in the
parameter space to probe its temperature profile,
because its equilibrium temperature straddles between two regimes where
either CO ($T > T_{\rm eq}$) or {\methane} ($T < T_{\rm eq}$) dominate
the carbon chemistry \citep[e.g.,][]{MosesEtal2013apjChemistryOfCOratios}, thus making the composition particularly
sensitive to the temperature profile.  We therefore adopted the
temperature as a main parameter to focus our exploration.

To model the atmospheric properties and theoretical spectra of
WASP-80\,b, we used the \textsc{Pyrat Bay}\footnote{https://github.com/pcubillos/pyratbay} open-source modeling
framework \citep{CubillosBlecic2021mnrasPyratBay}, We explored
potential physical scenarios under radiative and thermochemical
equilibrium (see details in Appendix \ref{sec:atm_modeling}),
parameterized by the planet's metallicity and C/O elemental ratio.
Certainly, given the equilibrium temperature of WASP-80\,b, processes
like photochemistry or transport-induced quenching can drive the
atmosphere out of chemical equilibrium
\citep[][]{Moses2014rsptaChemicalKinetics}.  Thus, we consider the
qualitative effects of disequilibrium chemistry on our models as well.

Figure~\ref{fig:equilibrium_physics} shows two examples of WASP-80\,b
models for a solar metallicity and C/O ratio, where we have shifted
the equilibrium temperature profile by $\pm 200$~K, and re-computed
the thermochemical equilibrium abundances and their respective
transmission spectra.
% Given the unknown albedos and efficiency of energy circulation on
% the planet, it is plausible.
The variation in temperature has a significant impact in both the
composition and the transmission spectra.  

For the higher-temperature model CO dominates the carbon chemistry,
with {\water} capturing the remaining---still largely
available---oxygen (thin dashed lines).  {\methane} is abundant in deeper layers, steadily decreasing with altitude.  In contrast, for the
lower-temperature model {\methane} dominates the carbon chemistry
(solid thick lines), {\water} dominates then the oxygen chemistry,
which leads to a much diminished CO abundance.
Spectroscopically, both {\water} and {\methane} should present
detectable absorption features in both models (having both broad
absorption bands across the 0.9--2.5 {\micron} spectrum), although the spectra are mainly shaped by {\water} bands at high temperatures and {\methane} at low temperatures, due to the different relative
abundances between these molecules (see bottom panel of
Fig.~\ref{fig:equilibrium_physics}).

The CO molecule, in contrast, only presents strong absorption in
narrow bands (1.6 and 2.5~{\microns}) and should only be detectable in
the spectrum of the higher-temperature model.  A conclusive detection
of CO would have favored the higher-temperature scenario, but this
is not the case.
%%%
The chain of reactions in the chemical network leads to
drastic changes in the abundances for most other trace species as
well, with {\carbdiox}, HCN, and {\acetylene} being more
abundant in the higher-temperature case by several orders of
magnitude.
%%%
The abundance of {\ammonia} is somewhat decoupled from the carbon and
oxygen chemistry, being more abundant at lower temperatures.
% being mainly determined by the net reaction: {\ammonia}
% $\rightleftarrows$ N$_2$ + 3H$_2$.
At their equilibrium abundances, neither {\ammonia} nor HCN would
present detectable features in the transmission spectra in both the low- and high-temperature models.
% The {\ammonia} molecule also has broad absorption bands at multiple
% wavelengths; however, in both cases its spectral features remain
% buried under the {\water} and {\methane} absorption lines.  Similarly,
% absorption bands of HCN (particularly at 1.05 and 1.5 {\microns}) are
% not detectable in the spectra due to the low HCN abundance.

%%% Disequilibrium chemistry:
The detection of {\ammonia} and HCN would thus
require to invoke disequilibrium chemistry processes that enhanced their abundances.  Specifically,
transport-induced quenching can drive the deep-interior abundance
throughout the atmosphere, with {\ammonia} and {\methane} being two major
species expected to be quenched.  The HCN abundance can be further
enhanced by a pseudo-equilibrium with the quenched {\methane} and
{\ammonia} and be produced photochemically by stellar ultraviolet
photons \citep[see][and references
therein]{Moses2014rsptaChemicalKinetics}.  With this in mind, both
modeled cases would show evidence of {\ammonia} absorption when
quenched from the lower layers.  However, HCN would only be detectable
on the higher-temperature case since its abundance at depth (from where it would be quenched) is $\sim$2 orders of magnitude larger than in the low-temperature case.  Thus, the detection
of HCN also points toward the higher-temperature case.

Finally, we also explored non-solar compositions.  In terms of the
global metallicity, {\carbdiox} is the molecule most strongly enhanced
with increasing metallicity, becoming detectable at metallicities
greater than
$\sim$10$\times$ solar. Thus, a confirmation of the tentative
detection of {\carbdiox} would point toward supersolar metallicities
in the atmosphere of WASP-80\,b.
Our results for sub-solar metallicity runs (0.1$\times$) are qualitatively similar to our solar ones.
In terms of C/O ratios we also
considered carbon-rich atmospheres, due to the strong impact on the
chemistry \citep[e.g.,][]{MadhusudhanEtal2011natWASP12batm,Madhusudhan2012apjCOratios,
  MosesEtal2013apjChemistryOfCOratios}.
%and possible previous suggestions of it \citep[][]{MadhusudhanEtal2011natWASP12batm}.  
We
found that elemental ratios of C/O > 1 are discouraged since the
{\water} abundance decreases, favoring then hydrocarbon species
like {\methane}, HCN, or {\acetylene}.
% While carbon-rich atmospheres have been heavily
% contested observatinally
% \citep[e.g.,][]{CrossfieldEtal2012apjWASP12b,
% SwainEtal2013icarWASP12bEmissionHST,
% KreidbergEtal2015apjWASP12bWFC3} and theoretically
% \citep[e.g.,][]{HellingEtal2014lifePlanetFormation,
% MordasiniEtal2016apjExoplanetFormation}.

The qualitative picture emerging from the comparison of the detected molecules with the atmospheric models we run provides tentative indications on WASP-80b's formation history. The metallicity of the host star WASP-80 is sub-solar \citep{Triaud2013}, whereas our observations suggest a subsolar-to-solar planetary metallicity. Within this range of possibilities, if the planet had a super-stellar metallicity,
combined with the solar C/O ratio estimation for WASP-80b,
our observations would
%so a solar metallicity for WASP-80b's atmosphere makes the planetary metallicity super-stellar.
%The compatibility of WASP-80b's C/O ratio with the solar value, when combined with the super-stellar metallicity,
favour formation scenarios where the giant planet accreted a significant mass of planetesimals while migrating through its native circumstellar disc \citep{Turrini2021a,Turrini2021b,Pacetti2022}. While the current data do not allow to draw stronger conclusions (e.g. on the extent of WASP-80b's migration), our NH3 detection makes WASP-80b a prime target for future efforts to quantify the abundances of C, O and N in its atmosphere and use their ratios to probe its formation history in more details \citep{Turrini2021a, Turrini2021b, Kolecki2021,Biazzo2022, Pacetti2022}. In particular, the present work will complement and help in the interpretation of \textit{JWST} spectra of WASP-80\,b whose transit and eclipse observations are planned through the GTO programs (PIDs 1177, 1185, and 1201).

\section{Conclusions and Future Perspectives}\label{sec:concl}
We report on the detection of multiple molecular species in the atmosphere of the warm Jupiter, WASP-80\,b, by analyzing nIR transmission spectra gathered with GIANO-B at TNG during four transits. We used two different statistical frameworks: the cross-correlation function technique and the likelihood approach. 
%While each molecule and each method are discussed in Sec. \ref{sec:detections}, 
We summarize the detections as follows: we report significant detections for H$_2$O, CH$_4$, NH$_3$ and HCN, tentative  detection for CO$_2$. We have inconclusive results for C$_2$H$_2$ and CO, whose presence in the WASP-80\,b atmosphere cannot be either firmly confirmed or excluded (see discussion in Sec. \ref{sec:detections}), but more observations and higher S/N data will help disentangling the nature of the signals (Figures \ref{fig:detectionssigma}, \ref{fig:detectionslike}).

%The former brought to the detection of water (with a significance of 7.5$\sigma$), methane (5$\sigma$), ammonia (5$\sigma$), hydrogen cyanide (4.1$\sigma$) and carbon monoxide (4.2$\sigma$). We found a tentative detection of acetylene, since it is detected by summing the first three nights with a significance of 3.7$\sigma$, but by adding the forth night a signal at v$_{\rm rest}$=63 km s$^{-1}$ become dominant, although the planetary signal is still evident (see Fig. \ref{fig:detectionssigma} and Table \ref{tab:detections}). The carbon dioxide detection is inconclusive, because the dominant peak in the CC map is at high values of K$_{\rm P}$, even though the planetary signal is evident with a significance difference of less than 1$\sigma$ from the prominent peak. By applying the likelihood approach we can firmly confirm the detections of H$_2$O, CH$_4$ and NH$_3$. We also obtained a tentative detection for C$_2$H$_2$ as for the CC framework, while no clear detection is found for CO. Finally, CO$_2$ showed a significant peak in the likelihood map at low K$_{\rm P}$ values, but it also showed a less significant signal at the nominal position (see Fig. \ref{fig:detectionslike}). 

The statistically robust detection of several species on WASP-80\,b
paves the way to estimate the chemical and physical conditions of the
planet's atmosphere.  Our initial exploration considering
radiative-thermochemical equilibrium models and the impact of
disequilibrium processes suggest that the atmosphere is compatible
with a solar composition ({\water} and {\methane} detections),
possibly affected by disequilibrium chemistry ({\ammonia} and HCN detections).  Future confirmation of the detection (or
non-detection) of CO will place strong constraints on the
temperature profile.
%, and therefore the dynamics and energetic regime of the planet.  
Likewise, a confirmation of {\carbdiox} will help to
constrain the atmospheric metallicity.

These results 
%together with those described in Guilluy et al. in prep., 
demonstrate for the first time that not only hot Jupiters \citep{Giacobbe2021}, but also warm giant planets present a rich chemistry in their atmosphere, breaking new ground in the study of exoplanetary atmospheres. With recent (e.g. CRIRES+, \textit{JWST}) and future upcoming (e.g., ELTs, \textit{ARIEL}) instruments, it will be possible to combine multiple wavelength bands and resolutions to derive accurate and precise molecular abundances as well as atmospheric elemental ratios (e.g. the C/O, N/O and C/N ratios) and metallicity, thus confirming/updating our qualitative estimates of solar C/O and metallicity under the assumption of thermo-chemical equilibrium, and our tentative constraints on WASP-80\,b's formation history. In particular, the combined information provided by the abundance ratios of elements with different volatility, like C, O and N revealed by our detections in WASP-80b's atmosphere, provides a direct window into the formation and migration history of giant planets \citep{Turrini2021a,Turrini2021b,Kolecki2021}. The improved estimates of the abundance of these elements achievable by such future facilities will therefore allow us to reconstruct the details of WASP-80b's formation history.

\begin{acknowledgments}
We acknowledge financial contributions from PRIN INAF 2019 and from the agreement ASI-INAF number 2018-16-HH. 
P. C. was funded by the Austrian Science Fund (FWF) Erwin Schroedinger Fellowship J4595-N. M.B. ackowledges support from from the UK Science and Technology Facilities Council (STFC) research grant ST/T000406/1. D.T. and E.S. acknowledge the support of the Italian National Institute of Astrophysics (INAF) through the INAF Main Stream project ``Ariel and the astrochemical link between circumstellar discs and planets'' (CUP: C54I19000700005) and ASI-INAF contract no. 2021-5-HH.0. VNa, IPa, GPi, and GSc acknowledge support from CHEOPS ASI-INAF agreement n. 2019-29-HH.0.
\end{acknowledgments}


\software{\\
\textsc{Pyrat Bay} \citep{CubillosBlecic2021mnrasPyratBay},
\textsc{repack} \citep{Cubillos2017apjRepack},
\textsc{TEA} \citep{BlecicEtal2016apsjTEA},
\textsc{Numpy} \citep{HarrisEtal2020natNumpy},
\textsc{SciPy} \citep{VirtanenEtal2020natmeScipy},
\textsc{sympy} \citep{MeurerEtal2017pjcsSYMPY},
\textsc{Matplotlib} \citep{Hunter2007ieeeMatplotlib},
\textsc{IPython} \citep{PerezGranger2007cseIPython},
and
\textsc{bibmanager} \citep{Cubillos2019zndoBibmanager}.
}

\vspace{5mm}
\facilities{TNG(GIANO-B)}  %% SWASP ? solo archivio
%\facilities{HST(STIS), Swift(XRT and UVOT), AAVSO, CTIO:1.3m,


%\software{astropy \citep{2013A&A...558A..33A,2018AJ....156..123A},  
%          Cloudy \citep{2013RMxAA..49..137F}, 
%          Source Extractor \citep{1996A&AS..117..393B}
%          }



\appendix

\section{Radiative-equilibrium Modeling}
\label{sec:atm_modeling}

To setup the \textsc{Pyrat Bay} modeling
framework to attain radiative and thermochemical
equilibrium, we solved the radiative-transfer equation in an iterative
approach, under the two-stream, plane-parallel, local-thermodynamic,
and hydrostatic approximations
\citep[following ][]{HengEtal2014apjsTwoStreamRT,
  MalikEtal2017ajHELIOS}.
%%%
To enforce chemical equilibrium, we calculated the compositions using
the thermochemical-equilibrium abundances code TEA
\citep{BlecicEtal2016apsjTEA} for input elemental
composition, temperature, and pressure profiles.  In this work we
modeled a chemical network containing H, He, C, N, O, Na, K,
H$_2$, {\water}, {\methane}, CO, {\carbdiox}, OH, {\acetylene},
C$_2$H$_4$, N$_2$, {\ammonia}, and HCN.
%%%
Simultaneously, to enforce radiative equilibrium the code updates the
temperature profile until the divergence of the upward--downward net
fluxes converges to a negligible value at each layer.  As boundary
conditions we imposed a 100~K blackbody internal radiative heat and an
incident stellar irradiation according to the properties of WASP-80,
assuming zero Bond albedo and full day--night energy redistribution.

We performed the radiative-transfer calculation over a fixed pressure
profile ranging from $100$ to $10^{-9}$~bar, and a wavelength grid
ranging from 0.3 to 30 {\microns} sampled with a 0.3 cm$^{-1}$
spacing, sufficient to encompass the bulk of the stellar and
planetary radiation (mainly in the optical and infrared,
respectively).
%%%
The radiative-transfer opacities include line lists for the most
relevant molecular species, i.e., CO, {\carbdiox}, and {\methane} from HITEMP
\citep{Rothman2010, LiEtal2015apjsCOlineList, Hargreaves2020} and of
{\water}, HCN, {\ammonia}, and {\acetylene} from ExoMol
\citep{Polyansky2018,
  Chubb2020,
  YurchenkoEtal2011mnrasNH3opacities,
  Harris2006, HarrisEtal2008mnrasExomolHCN,
  Coles2019}.
%%%
To handle the billion-sized ExoMol line lists, we employed the
\textsc{repack} algorithm \citep{Cubillos2017apjRepack} to extract
only the dominant transitions, reducing the number of transitions by a
factor of $\sim$100 without a significant impact on the resulting
opacities.
%%%
In addition to the molecular opacities, the \textsc{Pyrat Bay} code
included alkali resonance-line opacities for Na and K
\citep{BurrowsEtal2000apjBDspectra}; Rayleigh opacity for H, H$_2$,
and He \citep{Kurucz1970saorsAtlas}; and collision-induced absorption
for H$_2$--H$_2$ and H$_2$--He \citep{
  BorysowEtal2001jqsrtH2H2highT,
 Borysow2002jqsrtH2H2lowT, Richard2012}.

%\section{Appendix 2} \label{sec:pubcharge}
\section{Likelihood maps}
\label{app:B}

\begin{figure*}
\centering
\includegraphics[width=0.95\linewidth]{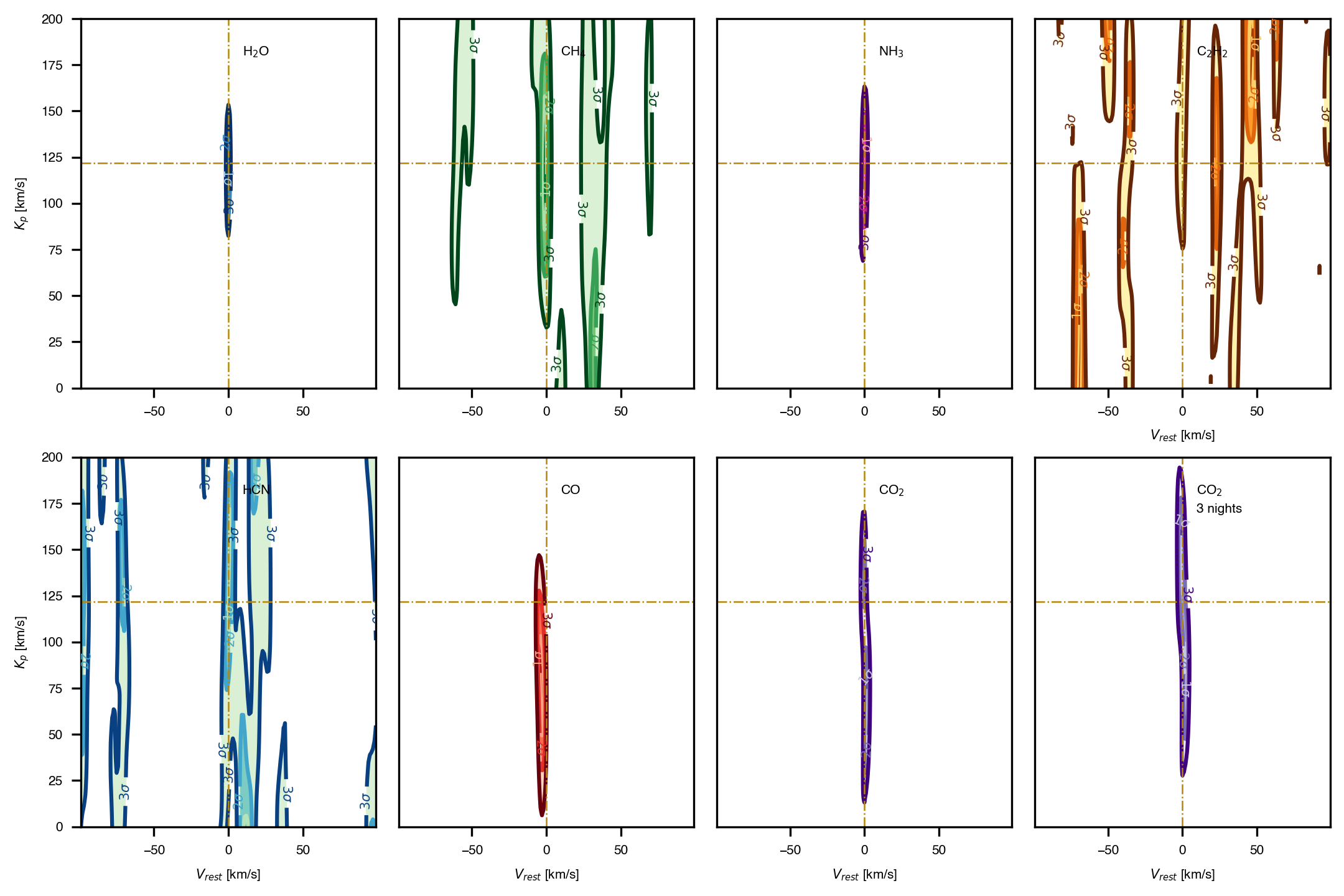}
\caption{\label{fig:detectionslike99} As Figure \ref{fig:detectionslike} but with v$_{rest}$ interval between [-99, 99]\,km\,s$^{-1}$.}
\end{figure*}


\bibliography{main_revised2}{}
\bibliographystyle{aasjournal}



\end{document}